\documentclass[10pt]{article}
\begin{document}
\begin{center}
\Huge Data-Adaptive Wavelets and
Multi-Scale SSA
\end{center}
\bigskip
\begin{center}
\Large Pascal
Yiou$^{\mbox{\ref{lmce},\ref{das}}}$,
Didier
Sornette$^{\mbox{\ref{ess},\ref{lpec}}}$ and
Michael
Ghil$^{\mbox{\ref{das}}}$\\
\end{center}
\begin{center}
Institute of Geophysics and Planetary Physics\\ University of
California Los Angeles at Los Angeles\\ Los Angeles, CA 90095-1567\label{igpp}
\end{center}

\bigskip
\begin{enumerate}

\item Permanent address\,: Laboratoire des Sciences du Climat et de l'Environnement, UMR
CEA--CNRS, CEA Saclay, 91191 Gif-sur-Yvette, France \label{lmce}

\item Also Department of Atmospheric Sciences, UCLA\label{das}

\item Also Department of Earth and Space Sciences, UCLA\label{ess}

\item Also Laboratoire de Physique de la Mati\`ere Condens\'ee, CNRS UMR 6622 and
Universit\'e de Nice-Sophia Antipolis, 06108 Nice Cedex 2,
France\label{lpec}

\end{enumerate}

\vfill

\begin{flushleft}
Physica D, submitted October 26, 1998.
\end{flushleft}

\newpage
\begin{abstract}
Using multi-scale ideas from wavelet analysis, we extend
singular-spectrum analysis (SSA) to the study of nonstationary time
series of length $N$ whose intermittency can give rise to the
divergence of their variance.
The wavelet transform is a kind of
local Fourier transform within a finite moving window
whose width $W$, proportional to the major period of interest, is varied to 
explore a broad range of such periods.
SSA, on the other hand, relies on the construction of 
the lag-covariance matrix ${\bf C}$ on $M$ lagged copies of the time
series over a fixed window width $W$ to
detect the regular part of the variability in that window in terms of
the minimal number of oscillatory components; here $W = M \Delta t$, with
$\Delta t$ the time step.
The proposed multi-scale SSA is a local SSA analysis within a
moving window of width $M\leq W \leq N$. Multi-scale SSA varies
$W$, while keeping a fixed $W/M$ ratio, and uses the
eigenvectors of the corresponding lag-covariance matrix ${\bf C}_M$ as a {\em
data-adaptive} wavelets; successive eigenvectors of ${\bf C}_M$
correspond approximately to successive derivatives of the first mother
wavelet in standard wavelet analysis. Multi-scale SSA thus solves
objectively the delicate problem of optimizing the analyzing wavelet in
the time-frequency domain, by a suitable localization of the signal's
covariance matrix. We present several examples of application to
synthetic signals with fractal or power-law behavior which mimic selected
features of certain climatic and geophysical time series. A real application is to the
Southern Oscillation index (SOI) monthly values for 1933--1996. Our methodology
highlights an abrupt periodicity shift in the SOI near 1960. This abrupt shift
between $4$ and $3$ years supports the Devil's staircase scenario for the El Ni\~no/Southern
Oscillation phenomenon.  \footnote{Preliminary results of this study were
presented at the XXII General Assembly of the European Geophysical Society,
Vienna, May 1997, and at the Fall Meeting of the American Geophysical Union, 
San Francisco, Dec. 1997.}

\end{abstract}

\tableofcontents

\newpage
\section{Introduction and motivation}
Intermittent time behavior is common in a large variety of systems in
the natural and socio-economic sciences.  Multiple regimes of behavior
and the transitions between them are ubiquitous in climate dynamics,
from the alternation of glacials and interglacials to that of
atmospheric blocking \cite{ghil-childress}.  River discharge and
rainfall similarly exhibit long periods of relative quiescence
interspersed with large floods and bursts \cite{Mandelbrot-1982}.  The
solid earth is characterized by even stronger intermittency that manifests itself
by long tails of earthquake-size distribution with strong
clustering of seismic activity in time and space
\cite{kagan-1994}. Even mantle convection is now believed to be
intensely intermittent, with internal boundary-layer
instability leading to ``avalanches'' that come in a large range of
sizes and follow a power-law distribution \cite{peltier-1996b}.
The number of contagious-disease incidences is also intermittent, and
has recently been documented to be distributed according to power laws
\cite{rhodesanderson-1996}.
The evolutionary path of species on earth with its ``punctuated
dynamics'' provides a paradigm of intermittent behavior
\cite{gould-1996}. Economic and financial time series have long been
recognized to have nonstationary variances, possibly described by
L\'evy processes \cite{Mandelbrot-1982}. 

These are only a few examples
among many that exhibit the ubiquity of intermittent behavior, and hence of
extreme events.
Additional examples range from volcanic eruptions, hurricanes and
tornadoes, landslides, avalanches, lightning strikes, and meteorite
and asteroid impacts, to the failure of engineering structures,
crashes in the stock market, social unrest leading to large-scale
strikes and upheaval, economic drawdowns on national and global
scales, regional power blackouts, and traffic gridlock. 

Many of these
problems involve nonstationary time series, while standard methods of
time-series analysis assume explicitly or implicitly
stationarity. Stationarity in the wide sense is defined by (i) the
existence of first (mean) and second (variance) moments of the time
series and (ii) their invariance with respect to time
translations. The failure of either assumption can lead to spurious
spectral-analysis results.

Wide-sense stationarity implies that the covariance operator of
a time series exists and is a function of the separation between epochs only,
{\it i.e.}, of ``lags.'' The spectral decomposition of the resulting
lag-covariance operator is connected to the power spectrum of the time
series through the Wiener-Khinchin theorem 
\cite{hannan-1960,percivalwalden-1993}. For time series in continuous time and of
infinite length, the corresponding eigenfunctions are always complex
exponentials or sine-and-cosine pairs, and one time series differs
from another solely through the associated eigenvalues.

For a time series of given, finite length --- in discrete time ---
singular-spectrum analysis (SSA) \cite{BK-1986,vg-1989} relies on
the Karhunen-Lo\`eve decomposition of an estimate of the covariance
matrix that is based on $M$ lagged copies of the time series. The
eigenvectors of this matrix are often called empirical orthogonal
functions, or EOFs and differ, in general, from sines and cosines
\cite{vyg-1992}. For the given time series, the EOFs form an optimal
basis that is orthonormal at zero lag; they permit one to
decompose the signal into its, possibly anharmonic, oscillatory
components and aperiodic ones. The data-adaptive character of the SSA
basis functions presents an advantage over the standard Fourier basis
of the classical Blackman-Tukey \cite{hannan-1960,percivalwalden-1993}
method: it allows one to capture a nonlinear oscillation by a single
pair of EOFs --- that maximizes the variance associated with the given
frequency --- rather than by a number of sine-and-cosine pairs
associated with that basic frequency and its harmonics.

SSA seems to be less sensitive than classical Fourier analysis to
deviations of a given time series from wide-sense stationarity assumptions. Trends are
captured quite successfully, most often by the leading one or two EOFs
of the given time series \cite{vyg-1992,ghilvaut-1991} and simple
discontinuities in the local value of an oscillation's frequency can
also be detected by SSA \cite{yiouetal-1995b,moronetal-1998}.

%
Wavelet analysis, on the other hand, has become a tool of reference
for intermittent, complex and self-similar signals, because it works
as a mathematical microscope that can focus on a specific part of the
signal to extract \emph{local} structures and singularities
\cite{meyer-1989,daub-1992,meyer-1993}.  The first step in the
definition of a wavelet transform is the choice of the analyzing
function, often called ``mother wavelet.'' 
Considerable work has
been devoted to finding mother wavelets that provide an orthogonal
expansion that is complete but not (highly) redundant. It is also
often desirable to concentrate the spectral energy in an optimal band
for the problem at hand, while keeping the localization property. 

A large number of analyzing wavelet functions have been
introduced to satisfy these often conflicting requirements, 
with their relative advantages and drawbacks. To provide
an optimal multi-scale decomposition of a signal, it may be desirable
to (A) automatically adjust the shape of the analyzing wavelet to the
signal rather than search through the extensive ``libraries'' of
mother wavelets (http://www.mathsoft.com/wavelets.html) 
and (B) modify this shape in time and scale, especially if the
data set's nonstationarity implies different structures as one slides
along time or scale. The intrinsic constraint of a unique mother wavelet does not
allow for this flexibility.

We introduce here a simple and powerful resolution to the limitations
of both types of methodology. The standard forms of the wavelet
transform and SSA are recalled in Secs. \ref{sec-wavelet} and
\ref{sec-ssa} respectively. The SSA method -- modified to use varying
windows with width $W$ proportional to the order $M$ of the
lag-covariance matrix ${\bf C}$ -- provides data-adaptive wavelet
transforms with analyzing functions given by the leading EOFs of the
matrix ${\bf C}$; it is described in Sec.~\ref{sec-msssa}. A set of
synthetic time series that exhibit intermittent and self-similar
properties, to wit, a Cantor set, a log-periodic process, and a
multiplicative noise, are introduced in Sec.~3.1; the details of their
construction appear in Appendices 7.1-7.4. We also
introduce a climatic time series, as a ``real-world'' test set. The
multi-scale SSA (MS-SSA) of Sec.~\ref{sec-msssa} is applied to these four test
sets in Secs.~\ref{sec-res-cantor} through \ref{sec-res-soi},
respectively. 

MS-SSA provides a new approach for testing
self-similarity and its breakdown by comparing the data-adaptive
wavelets at different scales, while it retains the usual scaling
of wavelet coefficients which is obtained from any wavelet transform
\cite{arneodoetal-1993}.
A summary of our results and comparison with complementary approaches
to address the requirements (A) and (B) above appear in
Sec.~\ref{sec-conclusion}. 

\section{Methodology}
\label{sec-method}

\subsection{Wavelet transform}
\label{sec-wavelet}
A wavelet transform requires the choice of an analyzing function
$\psi$, with general admissibility properties
\cite{meyer-1989,daub-1992,meyer-1993}, and with the more specific property of
time and frequency localization, \emph{i.e.}, $\psi$ and its Fourier
transform $\hat\psi$ must decay rapidly outside a given
interval. Functions based on a Gaussian, $f(x)=\exp( -x^2)$, possess
the localization property even if they do not verify the
admissibility condition that the integral of the 
mother wavelet vanish. This zero-mean condition ensures that knowledge of all
position- and scale-dependent wavelet coefficients allows one to retrieve the
initial function through an inversion formula (similar to a two-parameter
inverse Fourier transform) \cite{daub-1992}.

In the sequel, we shall follow \cite{arneodoetal-1993} and use a Gaussian wavelet and its 
first derivative for the sake of simplicity. The Gaussian itself does
not satisfy the admissibility condition, while  its first derivative does. The overall
statistical characterization of complex structures depends only weakly
on the choice of the mother wavelet \cite{arneodoetal-1993}
and we choose this simple example to illustrate its multi-scale
low-pass filter properties and contrast them to the proposed MS-SSA technique. 

A $\psi$-wavelet transform $W_\psi$ in continuous time and frequency
is simply a projection of a signal $X(t)$, $-\infty<t<\infty$, onto
$b$-translated and $a$-dilated versions of $\psi$:
\begin{eqnarray}
W_\psi(a,b) & = &\frac{1}{\sqrt{a}}
\int_{-\infty}^\infty X(t) \psi\left(\frac{t-b}{a}\right)
dt\label{eq-wave}\\
 & = &\frac{1}{\sqrt{a}} \int_{-\infty}^\infty
X(t+b) \psi\left(\frac{t}{a}\right)
dt.
\label{eq-wave2}
\end{eqnarray}
If most of $\psi$ is concentrated
between, $[-1,1]$ say (up to a rescaling), then Eq.~(\ref{eq-wave}) is
clearly an analysis of $X$ in the interval $[b-a,b+a]$, where the
integral is nonvanishing \emph{a priori}.

Using the successive derivatives $\psi^{(n)}$ of a given wavelet
$\psi$ in Eq.~(\ref{eq-wave}) is equivalent to a $\psi$-analysis (up
to a normalization factor) of the successive derivatives of the time
series $X$; this is easy to see through an integration by parts. It is
also easy to show, for a wide class of smooth analyzing functions $\psi$,
that the number of oscillations in
$\psi^{(n)}$ increases with $n$; this is true, in particular for our
Gaussian wavelet of choice.

The original signal (or a filtered version of it) can be reconstructed
from the family of wavelet transforms. Hence, for scale values $a$ in
an interval $I$, a reconstructed version $X_I$ of the signal $X(t)$ is:
\begin{equation}
X_I(t) = A_\psi \int_{a\in I}\int_{b=-\infty}^{\infty}
W_\psi(a,b) \psi\left(\frac{t-b}{a}\right) \frac{da db}{a^2};
\label{eq-wavrecons}
\end{equation}
$A_\psi$ is a normalization factor which only depends on the mother wavelet $\psi$.  This
formulation is essentially a bandpass filter of $X$ through $I$. If
$I$ is the positive real line, $I={\bf R}^+$, then $X_I=X$. 
Formula (\ref{eq-wavrecons}) applies when the mother wavelet is 
the derivative of the Gaussian, which
we use below. As
mentioned previously, the Gaussian $\psi(x)=\exp (-x^2/2)$ itself cannot be
used in this reconstruction formula, because it
does not satisfy $\int\psi(x) dx=0$. However, the forward transform
of Eq.~(\ref{eq-wave2}) is well-defined and provides nevertheless a useful
multi-scale decomposition of the signal. 

The counterpart of Eqs.~(\ref{eq-wave}) and
(\ref{eq-wavrecons}) for discrete time series can be obtained in a straightforward
way. Given a discrete signal $X_i=X(t_i)$,
$t_i=i \Delta t$, $0\leq i\leq N$, there are numerous ways of 
estimating Eq.~(\ref{eq-wave}) numerically
that avoid the problems of high-frequency noise and aliasing, while
minimizing computational cost 
\cite{meyer-1993,holschneider-1995}. Computationally efficient
algorithms based on the fast Fourier transform can be used
\cite{meyer-1993}, but we will only need here the simplest numerical
quadrature formulae to process Eq.~(\ref{eq-wave}).

\subsection{Singular spectrum analysis (SSA)}
\label{sec-ssa}
SSA has been described theoretically, for time-continuous signals
$X(t)$, by Vautard and Ghil \cite{vg-1989}. Since it is still less
familiar to a general readership than the wavelet transform, we
summarize here its form for time-discrete signals $\{X_i=X(i\Delta
t), ~0\leq i\leq N\}$, which is generally used in practice, as it will be
in the present study. Methodological and review papers include
\cite{vyg-1992,yiouetal-1995b,dettingeretal-1994,ghilyiou-1996}.  SSA
proceeds by finding the eigenvectors, called EOFs, of the
lag-covariance matrix ${\bf C}$ of the time series $\{X_i\}$ as above.
The true lag-covariance matrix ${\bf \tilde C}$ of the process that
generated $\{X_i: 1\leq i\leq N\}$ is estimated by a
lag-covariance matrix ${\bf C}_{M}$; assuming
stationarity of $X$ yields a Toeplitz structure
\begin{equation}
{\bf C}_{M,i}= \frac{1}{N-i} \sum_{j=1}^{N-i} X_j X_{j+i},
\end{equation}
where $i$ ($0\leq i\leq M-1$) is the diagonal index of the matrix, and
the time series $X$ is normalized to have zero mean.

The eigenvectors of this symmetric matrix are orthonormal and provide the Karhunen-Lo\`eve
basis for expanding the time series $X_i$ with respect to its $M$ lagged copies.
These EOFs ${\bf \rho}_k$ of ${\bf C}_M$ are sorted by decreasing order of the
associated eigenvalue $\lambda_k$, $0\leq \lambda_M\leq
\cdots\leq\lambda_2\leq\lambda_1$. The principal component (PC)
${\bf a}_k=\{a_{ki}:0\leq i\leq N-M\}$ is computed by projecting the time
series onto ${\bf \rho}_k=\{\rho_{kj}:0\leq j\leq M\}$:
\begin{equation}
a_{ki} = \sum_{j=1}^{M} X_{i+j-1} \rho_{kj}.
\label{eq-pc}
\end{equation}

For a given $k$, this equation is formally similar to Eq.~(\ref{eq-wave2}), in which the
integration interval is essentially restricted to the domain over
which the wavelet function is numerically nonnegligible. Furthermore, the EOFs $\rho_k$
have an oscillation property which resembles that of the derivatives $\psi^{(n)}$: the higher
the order $k$, the more sign changes $\rho_k$ has \cite{vg-1989,vyg-1992}. This
oscillation property is due to the formal analogy between the Toeplitz structure of the 
covariance matrix ${\bf C}_{M}$ and that of finite-difference discretizations of 
Sturm-Liouville problems on an interval $W = M \Delta t$ \cite{Courant}. In our case, $\Delta t$ is 
the sampling interval of the time series; in the discretized Sturm-Liouville problem, it
is the mesh size. In both cases, the true operator being approximated is positive definite
(or, more precisely, non-negative semi-definite), {\it i.e.}, its eigenvalues are all positive
(possibly one or more, but not all, being zero).

As in wavelet and Fourier analyses, time-dependent
reconstructions of the entire signal $\{X_i\}$ or of a suitable
filtered version thereof are possible from its PCs and EOFs. Hence, for a given
set of eigenelements $\cal K$, the corresponding
reconstructed component (RC) is \cite{vyg-1992}\,:
\begin{equation}
r_{{\cal K},i}=\frac{1}{M_i} \sum_{k\in {\cal K}}\sum_{j=1}^M a_{k,i-j}\rho_{kj},
\end{equation}
where $M_i$ is a scaling factor equal to $M$ when $i$ is sufficiently
far from the end points, and contains corrections when $i$ approaches
the interval's endpoints to within a distance $M\Delta t$.

\subsection{Multi-scale SSA (MS-SSA)}
\label{sec-msssa}
Systematic comparisons between 
SSA, the wavelet transform and other spectral-analysis methods have been carried out
in Ref. \cite{yiouetal-1995b,ghiltaricco-1997} (see, in particular, Table 1 in 
\cite{ghiltaricco-1997}). Further analogies between certain mathematical features
of SSA and wavelet analysis were mentioned in \cite{yiou-1994}. Table \ref{tab-paral} here
summarizes the most useful mathematical parallels between the two time-series analysis
methods; these parallels provide the basis for
extending global SSA analysis to a local one.  In SSA, the largest
scale at which the signal $X$ is analyzed in Eq.~(\ref{eq-pc}) is
approximately $N$ (the length of the time series), and the largest period is
$M$. As a consequence, the EOFs $\rho_k$ contains information from the
whole time series, as in the Fourier transform.

\begin{table}[h]
\begin{center}
\begin{tabular}{| l | c | c |}
\hline
Method & SSA & Wavelet transform\\
\hline
 Analyzing & EOF $\rho_k$ & Mother wavelet $\psi$\\
function & & \\
\hline
 & & \\
Basic facts & $\rho_k$
eigenvectors of $C_M$ & $\psi$ chosen \emph{a priori}\\
Decomposition &
$\displaystyle \sum_{t'=1}^M X(t+t') \rho_k(t')$ & $\displaystyle
\int X(t)\psi\left(\frac{t-b}{a}\right) dt$\\
Scale & $W=\alpha M$ & $a$\\
Epoch & $t$ & $b$\\
 & & \\
Average & $\rho_1$ & $\psi^{(0)}$\\
\& trend & & \\
Derivative & $\rho_2$ & $\psi^{(1)}$\\
\hline
\end{tabular}
\end{center}
\caption{Analogy between SSA and the wavelet transform.}
\label{tab-paral}
\end{table}

In order to define a local SSA, we propose to extend the SSA methodology by using a
time-frequency analysis within a running time window whose size 
$W$ is proportional to the order $M$ of the covariance matrix. Varying
$M$, and thus $W$ in proportion, we obtain a multi-scale
representation of the data.  We perform \emph{local} SSA on a time
series by sliding windows of length $W \leq N$, centered on times
$b=W/2,\dots,N-W/2$.  
When using this method, we assume that considerable information content
resides in the local variance structure and that the time series is locally
the sum of a trend, statistically significant variability, and noise.

\emph{A priori}, we can vary the two
scales $W$ and $M$ independently, as long as $W$ is larger
than $M$, $W/M\geq\alpha>1$, and $\alpha$ is large enough
\cite{vyg-1992}.  In the wavelet transform, however, for instance the
Mexican hat which corresponds to the second derivative of the Gaussian
function, the number of oscillations of the mother wavelet is fixed
and independent of the scale (width) of the analyzing wavelet. In
this spirit, we fix the ratio $W/M=\alpha$ and rely therewith on
the oscillation property of the EOFs described in Sec. 2.2 above 
to provide a fixed number of
zeroes for the data-adaptive ``wavelet'' $\rho_k$ of each local SSA
analysis; $\alpha=3$ in the calculations presented below and, as we
shall see, it will in fact suffice to use $k=1$ or $k=2$ on each $W$-interval.
This provides an analysis at a fixed scale $W$. Sampling a set of
$W$ values that follow a geometrical sequence, for instance in
powers of 2 or 3, provides a multi-scale analysis very similar to
the wavelet transform.

For a given position $b$ and fixed $W$, we obtain local EOFs, for
which PCs or RCs can be computed as described in
Sec~\ref{sec-ssa}. The EOFs are the direct analogs of analyzing
wavelet functions, as summarized in Table~\ref{tab-paral}.  Similarly to
successive analyzing wavelets (for instance the derivatives of the
Gaussian function), the number of EOF
oscillations increases roughly with order and the zeroes of $\rho_{k+1}$
separate those of $\rho_k$. This is consistent with the orthogonality of
distinct EOFs, with their being even or odd about the mid-point of the window,
and with their tending in certain cases to pairs of sines and cosines \cite{vg-1989} or
of Legendre polynomials \cite{Gibsonetal}. The first EOF thus corresponds approximately to an
analyzing wavelet function with a single extremum and no zero, for instance the
Gaussian. The second EOF is reminiscent of the first derivative of the
Gaussian and so on.

Then for each $b$ and each EOF $\rho_k$, it is possible to obtain
local PCs $a_k$ and RCs $r_k$. The $k$th PC at time $b$ is
\begin{equation}
a_{ki}^b = \sum_{j=1}^{M} X_{i+j-1} \rho_{kj}^b,
\end{equation}
and the corresponding RC is
\begin{equation}
r_{{k}i}^b=\frac{1}{M_i} \sum_{j=1}^M a_{k,i-j}^b \rho_{kj}^b,
\label{eq-rc-msssa}
\end{equation}
with $b-W/2 \leq i\leq b+W/2$.
The crucial difference between this local version and
 global SSA is that the RCs are
obtained here from local lag-covariance matrices. As $b$ varies from 
$W/2$ to $N-W/2$, this implies that the RCs will be truncated near the
edges of the time series.

We thus see that the local SSA method provides
simultaneous ``wavelet transforms'' of the data by a set of analyzing
wavelet functions, corresponding to the $M$ different EOFs of the
lag-covariance matrix. When $W$ (and thus $M$) is small, local SSA
provides a small-scale analysis of the signal with a few distinct
analyzing functions, namely a (sub)set of
EOFs indexed by $k$.  This is reasonable as there are not many
possible structures at scales that approach the sampling time
scale. On the other hand, at large scales, local SSA can also provide the
simultaneous analysis by many different analyzing mother wavelet functions, 
$\{\rho_k : 1 \leq k \leq M\}$, and thus
reflect the large possible complexity of the structures that can
develop over the entire time series.

The most important property of this local SSA analysis is that the
analyzing functions (analogous to the wavelets) are
\emph{data adaptive}: they are just the EOFs
of the lag-covariance matrix with time lags up to $M$, within a time window
of size $W = \alpha M$. In other words, the shape of these analyzing
functions is not imposed \emph{a priori}, like in a wavelet analysis,
but explicitly depends on the time series itself. This property is
particularly suitable in analyzing time series that exhibit dominant structures
which differ along the signal.  For instance, an oscillatory behavior
could be followed by white or colored noise and then by
deterministically intermittent behavior; this could indicate regime transitions 
that the system which generates the signal underwent while under
observation. If so, one would like to
have an analyzing wavelet which is adapted to each section of the signal. Our
data-adaptive scheme will definitely help follow such regime
transitions in time.

In addition, there is information to be gained from the shape of the
EOFs that are provided by the data.  Wavelet transforms are sometimes
used to test for self-similarity: one tests for the existence of a
power-law dependence of a wavelet coefficient at a given time as a
function of scale. If the power law is observed, one concludes on the
existence of a local singularity associated with such a law.  
This is useful in analyzing multifractal
structures characterized by a set of distinct singularities
\cite{arneodoetal-1993}.

The approach presented here provides an additional test. Indeed, if a
signal is locally self-similar, the shape of the EOFs must also be the
same, as we change scale ({\it i.e.}, $M$ and $W$).  We can thus carry
out an analysis at different scales, obtain the data-adaptive
analyzing functions and test for their self-similarity.  This
possibility arises naturally out of our multi-scale SSA (MS-SSA)
method and does not exist in any technique that assumes \emph{a priori} the
shape of the analyzing function, however carefully selected. We
present a few tests of MS-SSA below, and compare the results with
those of a standard wavelet analysis.

The interesting properties of MS-SSA come however at a significant
computational cost.  Indeed, the numerical cost for matrix
diagonalization far exceeds that of fast Fourier transforms or simple
algorithms for wavelet analysis, even when several wavelet functions
are used. In order to circumvent this problem and reduce the
computational cost, we carry out the local SSA analysis only for a
subset of time steps, $b=\beta \Delta t$. This sampling interval for
the analysis must be substantially smaller than the scale parameter
$W$, {\it i.e.} $\beta\ll W$. Then we use Eq.~(\ref{eq-rc-msssa}) for
$b\neq \beta\Delta t$ to
obtain the local RCs where the MS-SSA algorithm is not applied.  We
have checked that this interpolation procedure is very accurate when
compared to exhaustive local SSA analysis at each time step,
{\it i.e.} computing $r_{kb}^b$ in Eq.~(\ref{eq-rc-msssa}) for each $b$ and
hence discarding its values at epochs surrounding $b$.  This procedure
is based on the relative robustness of the EOFs, which is also used in SSA-based
time-series prediction \cite{vyg-1992,KeppenneGhil92,GhilJiang98}.

\section{Data sets}
\label{sec-data}

\subsection{Synthetic data sets}

We would like to investigate the behavior of MS-SSA on time series
that are self-similar or fractal, as they pose {\it a priori} the biggest challenge to 
techniques based on well-behaved finite covariances. 
As already mentioned, such features
are ubiquitous in the geosciences \cite{dubrulleetal-1997}.

The first time series $\tilde P_1$ that we shall investigate was obtained by
using an Iterated Function System (IFS) \cite{barnsley-1993} to approximate
a simple triadic set. The procedure that
we have used is described in Appendix 7.1 and the resulting time-series is shown
in Fig. 1. A closely related time series
$P_1$ is obtained from the characteristics of the exact triadic Cantor set.
Its construction is described in Appendix 7.2.
In the ``perfect'' construction scheme of $P_1$, we anticipate large
``resonances'' in the lag-covariance matrix that should be smoothed
out by the noise present in the construction of $\tilde P_1$. 
Finally, a multiple-rule Cantor set is generated in Appendix 7.2 to yield the
time-series ${\hat P}_{1}$.

\begin{figure}[h]
\caption{A 1000-point realization of the $\tilde P_1$ process
(top) and its blow-up (bottom).}
\label{fig-P1}
\end{figure}

The second process is a log-periodic signal with a square-root
singularity. In the context of critical phenomena, the log-periodicity
corresponds to the existence of an imaginary part in the critical
exponent, and is associated with a discrete scale invariance
\cite{sornette-1998}, {\it i.e.} to the invariance of the system or of its properties
only under magnifications that are integer powers of a fundamental ratio.
Initially, complex exponents were proposed as
formal solutions of renormalization-group equations in the seventies
\cite{Jona,Nauen,Nieme}. In the eighties, they have been shown to emerge in
 various physical problems that arise in discrete hierarchical
systems \cite{DIL,Fournier}. 

Recently, it has been realized that discrete
scale invariance and its associated complex exponents and log-periodicity may appear
``spontaneously,'' without the need for a pre-existing hierarchy, as a result of
dynamical processes. Such behavior has been found in models and experimental 
data from irreversible growth
processes, rupture in heterogeneous systems, earthquakes, percolation
and even in financial crashes \cite{crash} (see \cite{sornette-1998} for a review,
 and references therein).  
Another source of time series with log-periodic behavior is provided
by Boolean delay equations, whose solutions are Boolean-valued
processes in continuous time \cite{ghilmulh-1985}. These equations have
been used to model paleoclimatic variability 
\cite{DarbyMysak93,Ghiletal87paleo}, interdecadal \cite{Wrightetal90paleo} and most recently,
seasonal-to-interannual \cite{Saunders} climate change, although not all of these
applications exhibit the behavior in question. In the case of Boolean delay equations,
the log-periodicity exhibits a period and amplitude that increase as
$t\to t_c=+\infty$ \cite{ghilmulh-1985} and their treatment by MS-SSA is left for
subsequent study. 

The log-periodicity we concern ourselves with here
is reflected in oscillations of decreasing period and
amplitude that are superimposed on a power-law behavior and culminate
at a singularity $t_c$ in finite time. At the critical point $t_c$,
the instantaneous frequency of the oscillation becomes infinite, but
its amplitude vanishes. Again, a global spectral analysis method would
fail to represent this signal in a satisfactory manner since its
nonstationarity is characterized by a continuously varying period and
a nonlinear trend. The log-periodic time series $P_2$ is constructed
according to the procedure given in Appendix 7.3 and is shown in Fig. 2.

\begin{figure}[h]
\caption{Log-periodic time series $P_2$.}
\label{fig-lopger}
\end{figure}

While the first two processes $(P_1, {\tilde P}_1)$ and $P_2$ are purely deterministic, 
the third process we have studied still presents self-similar properties,
albeit with a large stochastic component. Its
self-similarity stems from its multiplicative-noise nature.  The process $P_3$ has the
property of being statistically stable under affine transformations
\cite{sornettecont-1997,sornetteknopoff-1998}. A 512-point realization of $P_{3i}$, for
$1\leq i\leq 512$, is shown in Fig.~\ref{fig-p3} and a histogram is
provided to illustrate the power-law behavior and its long tail.

\begin{figure}[h]
\caption{Multiplicative-noise time series $P_3$.}
\label{fig-p3}
\end{figure}

\subsection{Climatic data set}

In addition to the synthetic data sets that possess well-understood
properties, we chose to analyze a ``real-world'' climatic time series,
the Southern Oscillation Index (SOI).  This time series illustrates well another
type of behavior often encountered in the geosciences, viz., the superposition
of quasi-periodic and aperiodic behavior \cite{GhilJiang98,ghiletal-1991}.
SOI is a climatic index
connected with the recurring El Ni\~no/Southern Oscillation (ENSO)
conditions in the tropical Pacific; it is essentially the monthly mean
difference in sea-level pressure between Darwin, Australia, and
Tahiti (Fig.~\ref{fig-soi}). An anomalously negative value of this
index indicates a warm ENSO event (El Ni\~no), while a highly positive value is
associated with a cold event (La Ni\~na).
SOI has been used in numerous
studies for understanding the dynamics of ENSO \cite{jiangetal-1995} and improving its
prediction \cite{KeppenneGhil92,GhilJiang98}. 

In the data set we use here, the annual cycle was removed
and the time series was normalized by its variance. The time interval
considered goes from January 1933 to December 1996, during which very
few observations are missing at either station.

\begin{figure}[h]
\caption{SOI variations between 1933 and 1996.}
\label{fig-soi}
\end{figure}

ENSO models have shown that, in the Tropical Pacific, 
phase locking of the coupled ocean-atmosphere system's self-sustained
oscillation to the annual cycle leads to a Devil's staircase
\cite{jinetal-1994}. This behavior has been confirmed by studies of
sea-surface temperature data sets using both classical (global) SSA
and (fixed-analyzing function) wavelet methods.
Such studies have been performed on the sea-level pressure record at Darwin only
\cite{wangwang-1996}, which spans a longer time interval, using
wavelet and waveform analysis, and Tropical Pacific sea-surface temperatures
 with global SSA \cite{moronetal-1998,Rasmusson,jiangetal-1995}. Two of these
 analyses have shown a fairly sudden shift in
the characteristic period of ENSO's low-frequency component that occurred
in the 1960s, with a longer period before and a shorter one prevailing since
\cite{moronetal-1998,wangwang-1996}.

\section{Numerical results}
\label{sec-res}

\subsection{Cantor-set experiments}
\label{sec-res-cantor}

\subsubsection{SSA analysis}

Global methods of time-series analysis will clearly miss the
multi-scale feature of the IFS construct $\tilde P_1(t)$, and
essentially see one or more nearly periodic components that arise
from a smoothed version of the exact Cantor set $P_1(t)$. A
Monte-Carlo SSA \cite{ghilyiou-1996,allensmith-1995} of $\tilde
P_1(t)$ with a lag of $M=40$ shows two frequencies that appear as pairs of
eigenvalues (0.12 and 0.25 cycles per time unit) emerging high above a
red-noise-like background; two additional eigenvalues stand slightly
above the red-noise background at frequencies 0.006 and 0.06 cycles
per time unit (Fig.~\ref{fig-mcssa-p1}). The pair at frequency 0.25 (period
4) represents the small-scale oscillations in $\tilde P_1$; the
smallest possible scale at this step of the IFS
is 2, but it is not systematically reached throughout the interval of interest,
so that the
smallest scale which is always present is 4 (see
Fig.~\ref{fig-P1}b). The pair at frequency 0.12 cycles per unit
(0.12 cpu; period 8.33) accounts for the grouping of two smaller-scale
oscillations. Finally, the pair near 0.09 cpu (period 11.1; not shown) appears to be
closely associated with the 0.12 cpu pair; hence the latter two pairs should be viewed as a
quadruplet. The first eigenvalue has an associated period ($156$) larger
than the lag window $M=40$, so that it has no statistical significance; visually,
it corresponds to roughly half the length of the largest-scale alternation between
quiescence and oscillations ($1/6$ of the entire interval).

\begin{figure}[h]
\caption{Monte-Carlo SSA of $\tilde P_1$.}
\label{fig-mcssa-p1}
\end{figure}
SSA enhances one characteristic scale (which depends on $M$) and
smears out the faster ones. On the other hand, the RCs of the low-order EOFs
exhibit mixtures of fast and slow oscillations. This is so because all the oscillations have
nearly the same variance by construction (Fig.~\ref{fig-rcssa-p1}): the
amplitude of small-scale oscillations is exactly the same as the one
of large-scale oscillations, and so the variance only depends on the distribution
of the oscillations over the interval.
The fast oscillations can thus be viewed as spurious because
they do not provide information on the way the data set was
constructed. Other global methods 
(Maximum Entropy: \cite{yiouetal-1995b,ghilyiou-1996,ghiltaricco-1997,childers-1978}; 
Multi-Taper: \cite{yiouetal-1995b,ghilyiou-1996,ghiltaricco-1997,thomson-1982}) do
not provide more relevant information on $\tilde P_1(t)$ either (not shown).

\begin{figure}[h]
\caption{Global RCs of $\tilde P_1$.}
\label{fig-rcssa-p1}
\end{figure}

\subsubsection{Wavelet transforms}

We next performed wavelet transforms of $\tilde P_1$ with the Gaussian
analyzing function $\psi(x)=\exp (-x^2/2)$ and its derivative
$\psi'$. Geometric scale increments $a=3^n$, $n=0,\dots,6$, were
used. The choice of an increment of $3$ is suggested by the triadic
structure of the Cantor set; other choices (e.g., $a=2^k$) clearly
cannot provide the same optimal scale decomposition (see below). More generally,
when the data set contains a discrete scale invariance with respect to
dilations, the use of the preferred scaling ratio --- such as the
ratio $3$ in the Cantor set studied here --- is optimal
\cite{arneodoetal-1993,dubrulleetal-1997}. The wavelet analyses of
$\tilde P_1$ are shown in Fig.~\ref{fig-wavt-P1}.  The triadic and
self-similar structure of the Cantor set is clarified, as each scale
picks up one step of the construction process.
\begin{figure}[h]
\caption{Wavelet transforms of $\tilde P_1$ with scale increments of 3.}
\label{fig-wavt-P1}
\end{figure}

A wavelet analysis with $\psi'$ as the mother wavelet (right column of Fig. 7),
rather than $\psi$ (left column), reveals further aspects of the process's
singularities. The triadic
construction of $\tilde P_1$ is also apparent in this decomposition, albeit not as
clearly as with $\psi$. The wavelet transforms of $\tilde P_1$ in
Fig.~\ref{fig-wavt-P1} do not differ significantly from those of its
``exact'' counterpart $P_1$ (not shown).

\subsubsection{MS-SSA analysis}

\noindent $\bullet$ {\it RC analysis}. We performed 
MS-SSA analyses of $\tilde P_1$ with scales
$W$ varying with geometric increments, $W = 2 \cdot 3^n$,
and a ratio $W/M = 3$. We kept the first two EOFs for
each scale. The first RCs of $\tilde P_1$ reveal very well
the triadic structure of the Cantor set (left column of Fig.~\ref{fig-msssa1-p1}), 
as in the Gaussian-wavelet analysis.

\begin{figure}[h]
\caption{RCs 1 and 2 of $\tilde P_1$ for scale increments of 3.}
\label{fig-msssa1-p1}
\end{figure}

Unlike the wavelet analysis, MS-SSA restricted to EOF-1 reproduces
almost exactly the steps taken to generate $\tilde P_1$, {\it i.e.}, the
divisions do not overlap, even though parts of the reconstruction are
lost due to side effects. Thus, in the case of $\tilde
P_1$, the triadic structure seems to be more faithfully captured by
this method. A slight drawback, on the other hand, is that SSA
exhibits a mild ``Gibbs effect,'' {\it i.e.}, an overshoot at certain
discontinuities; it does not guarantee, therewith, the positiveness of
the signal reconstruction which is ensured by a wavelet analysis
with a positive function, like the Gaussian (but not its derivative: see right
column of Fig. 7).

The RCs corresponding to EOF-2 (right column of
Fig.~\ref{fig-msssa1-p1}) provide an analysis of the Cantor set's
``derivative,'' as with the $\psi'$-wavelet transform. Hence, EOF-2
is analogous to $\psi'$ (see Table~1 and ref.\cite{Gibsonetal}), and the
resulting RCs yield an analysis of the derivative of ${\tilde P}_1$ (not
shown).

\vskip 0.5cm
\noindent $\bullet$ {\it EOF analysis}.
As anticipated in Section~\ref{sec-method}, we expect the EOFs
$\rho_k^{(W)}$ to be similar for fixed order $k$ and on
different scales $W$, due to the choice of the scaling ratio $3$
being equal to that used in constructing the time series.
Figure~\ref{fig-eof-p1} shows EOFs 1 through 4 of $\tilde P_1$, for
scales $W$ in a fixed ratio of 3 between 9 and 243 (a scale of 3
implies $M=1$, for which the analysis is trivial), and at a
location $b=333$, situated at one third of the time series.
The EOFs are scaled by ${3}^{n/2}$, so that their amplitudes match. We
notice a near-convergence in the shape of EOFs 1--3 as the scale
increases; this indicates that the time series is
self-similar with scale increments of 3. EOF 4, on the other hand, needs larger
scales (and more data) in order to converge.  

\begin{figure}[h]
\caption{EOFs $1-4$ of $\tilde P_1$ at $b=333$, with a scaling of $3^{n/2}$.}
\label{fig-eof-p1}
\end{figure}

In contrast, we show in Fig.~\ref{fig-eof-p1-2k}
the same EOFs 1-4 of $\tilde P_1$, for scales $W$ in a
fixed ratio of 2, between 16 and 256, at the same position $b=333$
along the time series. There is practically no convergence with $W$ 
for fixed $k$, reflecting the mismatch between the scale factor 2
of the multiscale analysis and the preferred scale factor 3 of the
time series. This suggests a novel method for detecting the existence
of a preferred scaling factor $\lambda$ in a time series by optimizing
the scale ratio $l$ in MS-SSA: only when $l \approx \lambda^m$, where
$m$ is an integer, will the convergence of the EOFs $\rho_k^{(W)}
\to \rho_k^{(\infty)}$ be good as $W \to \infty$.

\begin{figure}[h]
\caption{EOFs $1-4$ of $\tilde P_1$ at $b=333$, with a scaling of $2^{n/2}$.}
\label{fig-eof-p1-2k}
\end{figure}

This phenomenon reflects a ``scale-locking''
between the analyzing (multi)scale factor and the intrinsic scale
ratio(s) present in the time series.  This provides a scaling test that
conventional wavelet analysis cannot offer, due to the \emph{a priori}
choice of $\psi$. In MS-SSA, the functional-shape constraint is relaxed
and new information on the signal can thus be obtained.

\vskip 0.5cm
\noindent $\bullet$ {\it Resonance}.
An interesting phenomenon appears when applying MS-SSA to the
``perfect'' Cantor set $P_1(t)$. The local lag-covariance matrices
seem to exhibit resonances between scales so that a given ``slow''
scale contains information on faster scales and hence exhibits
oscillations that blur the triadic decomposition (not shown). This was not seen in
the MS-SSA of the ``approximate'' time series $\tilde P_1(t)$, which
contains noise in both time and scale that smoothes out such
resonances and leads to a clear reconstruction of the triadic
structure.  The presence of a small amount of noise often increases
the robustness, and hence ease of identification, of a phenomenon's
main features. Well-known instances are the
randomization method in general probability theory  \cite{Feller}
and the
addition of process or observation noise in sequential estimation
theory \cite{ghilrizzoli-1991}.

\vskip 0.5cm
\noindent $\bullet$ {\it Change of geometrical structure along the scale axis}.
 MS-SSA analysis of the multiple-rule Cantor set $\hat P_1$
(Fig.~\ref{fig-msssa-p1mlt}) shows that the method can capture the difference
in structure beyond a threshold scale, even though the transition may
be smoothed out by the two construction procedures. The RCs do not
give as clearcut a triadic reconstruction of this Cantor set as for the
single-rule set $\tilde P_1$ in Fig.~\ref{fig-msssa1-p1}.
But, in spite of the relatively small data set we used (only 729 points),
we can still detect in Fig.~\ref{fig-msssa-p1mlt} the
difference between the use of rule (\ref{rule1}) for the larger scales and
rule (\ref{rule3}) for the smaller ones at the passage between the scales $W=9$ and
$W=27$.

\begin{figure}[h]
\caption{RCs 1 and 2 of $\hat P_1$ for scale increments of 3.}
\label{fig-msssa-p1mlt}
\end{figure}

\vskip 0.5cm
\noindent $\bullet$ {\it Reconstruction process by summing over scales for 
a single EOF}. By construction, there is a finite number $M$ of
EOFs for each scale $W$ in MS-SSA. The EOFs form an
orthonormal basis and hence the local decomposition of the time series
occurs on a finite number $M$ of modes, at each scale. Thus, for a
given scale $W$, the sum of the corresponding RCs reproduces
 the original time series in each window. In wavelet language, this corresponds
to reconstructing a signal by summing, at a fixed scale $a$ (see Table 1), 
over a finite number of wavelet
transforms; each of these transforms
uses a different mother wavelet -- for instance, the successive derivatives $\psi^{(n)},
0 \leq n \leq M-1$, of the same mother wavelet $\psi = \psi^{(0)}$ -- 
such that the finite set
of mother wavelets form a (not necessarily orthogonal, but still nondegenerate)
basis. This, however, does not correspond to the
usual reconstruction formula (\ref{eq-wavrecons}).

In {\it continuous} wavelet analysis, the reconstruction is usually performed by summing over
the scales for a fixed mother wavelet, according to Eq. (\ref{eq-wavrecons}),
often called a ``resolution of the identity (operator)'' on admissible functions \cite{daub-1992}.
For discrete wavelet transforms, there does not exist, in general, a ``resolution
of identity'' formula analogous to (\ref{eq-wavrecons}): 
the expansion of a function over a discrete
wavelet basis is not orthogonal in general and special iterative methods must be
developed to derive an inversion formula (see Chapter 3 in \cite{daub-1992}). 

Likewise, for our MS-SSA method, the expansion over discrete scales of 
a function with a fixed EOF order is not orthogonal in general. The
MS-SSA approach is thus more closely related to the theory of ``frames'' \cite{daub-1992}.
It is easy to show from the 
definition of a frame that the family of EOFs
${\bf \rho}_k^{W}$ of given order $k$ for discrete scales $W_n=a^n$ with $n$ integer
constitutes such a frame. This yields an algorithm to reconstruct the
signal from the discrete set of wavelet transforms, in our case the RCs of same order
$k$ at different scales, for special choices of the scale $W$ and translation 
parameter $b$ \cite{daub-1992}\,:
\begin{equation}
X_i = \sum_{m} \sum_{j=1}^M {\bf \hat \rho}_{k,j}^{W_m} a_{k,i-j}^{W_m}~;
\label{ghjhjwlk}
\end{equation}
here the sum over $m$ corresponds to the different scales, and $a_{k,i-j}^{W_m}$ is given by
\begin{equation}
a_{k,i-j}^{W_m} = \sum_{l=1}^M {\bf \rho}_{k,l}^{W_m} ~X_{i-j+l-1}  ~,
\end{equation}
while ${\bf \hat \rho}_{k,j}^{W_m}$ is a vector related to the EOF ${\bf \rho}_{k,j}^{W_m}$
through an iterative algorithm \cite{daub-1992}. 

The inversion can thus be formulated
in principle, but in practice it is quite laborious. Emphasizing the practical aspect, we propose
an approximate inversion formula similar to (\ref{ghjhjwlk}) but with 
${\bf \hat \rho}_{k,j}^{W_m}$ replaced by ${\bf \rho}_{k,j}^{W_m}$\,:
\begin{equation}
X_i = \sum_{m} \sum_{j=1}^M {1 \over M} {\bf \rho}_{k,j}^{W_m} a_{k,i-j}^{W_m}~.
\label{dsfgsg}
\end{equation}
This formula is tested in Fig. \ref{fig-inversion-p1}. The top panel shows the time
series $P_1$. The second panel shows the RC using EOF-2 ($k=2$) for the single scale $W=9$. This
corresponds to formula (\ref{dsfgsg}) with a single value of $k=2$, while $a=3$ and thus
$W_2 = 9$.
The third, fourth and fifth panels show the sum (\ref{dsfgsg}) with two, three and four terms;
these terms correspond to $W_2 = 9$, $W_3 = 27$, $W_4 = 81$ and $W_5 = 243$, while the order
of the EOF $\rho_k^{W_m}$ is always $k=2$.

\begin{figure}[h]
\caption{Reconstruction of $P_1$ using Eq. (\ref{dsfgsg}) for scale increments of 3.}
\label{fig-inversion-p1}
\end{figure}

\subsection{A log-periodic process}
\label{sec-res-logper}
In this subsection, we focus on the detection and quantification of
self-similar structures in a log-periodic process. In the case of
Eq.~(\ref{eq-logper}), we expect discrete self-similarity near
$t=t_c$, with a preferred scaling of $\lambda=2$.  Indeed, in
Fig.~\ref{fig-lopger} the distance between successive maxima of the
oscillations decreases geometrically with a scaling of about $2$, and
thus converges to zero as $t$ approaches $t_c$. In the case of $P_2$,
a straightforward application of
MS-SSA (with EOFs 1 and 2) and of the wavelet transform (with $\psi$ and $\psi'$) give extremely 
similar results; they both act as low-pass
filters with geometrically increasing width (not shown) and hence do not provide
a satisfactory reconstruction of the signal.

We analyzed therefore $P_2(t)$ by MS-SSA with scaling increments of 2 but placing
ourselves near $t_c=100$ in the sense that the local windows always
end at $t_c$, and their midpoint moves to the left. The first
four EOFs, shown in Fig.~\ref{fig-eof-p2}, clearly converge rapidly as
the size of the window --- and thus the scale analyzed ---
increases. If we choose another ratio, say 3, to scale the windows no
such convergence is observed (not shown), as was the case in comparing
Fig.~\ref{fig-eof-p1} with Fig.~\ref{fig-eof-p1-2k} for the Cantor
set. This sensitivity of MS-SSA's EOFs to the correct
scaling factor suggests a new way to characterize log-periodicity.

\begin{figure}[h]
\caption{EOFs 1--4 of $P_2$ near $t_c$, with $2^n$ scaling.}
\label{fig-eof-p2}
\end{figure}

EOFs 1 and 3 capture the local mean of the signal and scale
the period of its oscillation, respectively, having no zero and two
zeroes (see Fig.~13). RC-2 describes best the local fast
oscillations near the critical point $t=t_c$, while RC-1 at different
scales captures well the overall behavior of the time series in
Fig.~\ref{fig-lopger} (see Fig.~\ref{fig-msssa-p2}).

\begin{figure}[h]
\caption{RCs 1 and 2 of $P_2$ for scale increments of 2.}
\label{fig-msssa-p2}
\end{figure}

We look next at what happens near $t=0$, {\it i.e.}, we use windows starting
at the beginning of the time series.  The convergence of EOFs 2--4 is
still excellent as $W$ increases (not shown). Only EOF-1, which
is sensitive to the overall behavior of the time series, exhibits
slower convergence with $W$, due to the lack of exact
self-similarity of $P_2(t)$ near the origin. This robust, albeit
imperfect, convergence of its EOFs for log-periodic time series makes
the MS-SSA approach potentially useful for the analysis of real
log-periodic data, in the presence of noise and truncated singularities. 
The application of this technique to real data on earthquakes 
\cite{SSearthquake,Johansen}  and financial  \cite{crash}
time series will be investigated separately.


\subsection{Multiplicative noise}
\label{sec-res-multnoise}
As expected, classical global methods which assume the existence of
finite variance do not give a particularly good insight into the
process $P_3(t)$. Monte-Carlo SSA yields a periodicity that is
statistically significant above 95\%; the actual period, though, lies
between 3 and 4 cycles and fluctuates from
realization to realization (not shown).
The time series cannot be otherwise distinguished from
red noise; this is obviously wrong and can be
explained by the lack of stability in the global variance estimate.

The global RCs (Fig.~\ref{fig-rcssa-p3}) provide a sensible
scale-dependent reconstruction for the slower scales: the
characteristic scale halves from RC-1 to RCs 2--3, on to RC-4 and, again to RCs
5--8, but the reconstructions become noisier
as the order increases further (not shown). The total reconstruction of
$P_3(t)$ from the eight leading RCs (Fig.~\ref{fig-rcssa-p3}e) is an
excellent coarse-scale representation of $P_3(t)$, as depicted in
Fig.~\ref{fig-p3}.

\begin{figure}[h]
\caption{Global RCs of $P_3(t)$.}
\label{fig-rcssa-p3}
\end{figure}

Next, wavelet and MS-SSA analyses were both performed with geometric
scale increments of 2, while the latter used a ratio $W/M=3$ as
in Secs.~\ref{sec-res-cantor} and \ref{sec-res-logper}.
A wavelet analysis using the Gaussian mother wavelet (Fig.~16, left column)
progressively smoothes the details of the time
series as the scale parameter $a$ increases. The analysis reveals
the natural self-similarity of $P_3$, as the analysis at scale $a=32$
can be normalized to match the first half of the analysis at
scale $a=8$ (Fig.~\ref{fig-wavlt-p3}).
In turn, the analysis at scale $a=8$ can be normalized to fit the
analysis at scale $a=2$ in the same way. Thus, we can infer that scale increments of
about $4$ are sufficient for the analysis of $P_3$.

\begin{figure}[h]
\caption{Wavelet analysis of $P_3$.}
\label{fig-wavlt-p3}
\end{figure}

MS-SSA using EOF-1 (Fig.~\ref{fig-msssa-p3}, left column)
behaves in a similar way, with a progressive smoothing of the time
series and the same scaling result.
RC-2 captures the oscillating variability (right column of Fig.~\ref{fig-msssa-p3})
around EOF-1, as does $\psi'$
with respect to $\psi$ (right column of Fig.~16).
In this case, the parallel between the
behavior of wavelet analysis and MS-SSA is striking.  As we saw in Fig.~15e, 
the reconstruction of $P_3$ can also be completed with a finite
sum of EOFs, for a fixed scale of analysis, whether global of not.


\begin{figure}[h]
\caption{RCs 1 and 2 of $P_3$ with scale increments of $2$.}
\label{fig-msssa-p3}
\end{figure}

In contrast to the results found in Fig.~\ref{fig-eof-p1} for the
Cantor set $\tilde P_1$, while the EOFs of the same order have a
similar shape across scales, they cannot be exactly superimposed (not
shown). This simply reflects the fact that the time series $P_3$ is not exactly
self-similar with a preferred scaling ratio of $2$.  In fact, no choice
of a single scaling ratio provides a good convergence of the EOFs for fixed $k$: $P_3$ is not
self-similar but rather exhibits a more complex scaling structure,
that is approximately multifractal \cite{sornetteknopoff-1998}.  The shape of
the EOFs (not shown) is in fact quite similar to that of the Gaussian
wavelets. 
The data-adaptive virtues of the EOFs are
illustrated in Fig.~\ref{fig-mesheof-p3}, where the shape of the local
EOF-1 follows closely the variations of $P_3$. 

\begin{figure}[h]
\caption{The evolution in time of the local EOF-1 for $P_3$.}
\label{fig-mesheof-p3}
\end{figure}

On this data set, the two methods under consideration perform in a rather similar way. The advantage
of MS-SSA is that it fully uses the information contained in
the time series in order to compute analyzing functions, and that the
reconstruction is intrinsically complete, given a small set of EOFs, which
is not the case for wavelets.

\subsection{Southern Oscillation Index}
\label{sec-res-soi}
We performed MS-SSA on the monthly SOI data for the years 1933-1996 
(see Section 3.2). The parameters were, as before, $W/M=3$ and geometric scale
increments of $2$. In addition to the previous analyses, we also
computed an ``instantaneous'' frequency for each local
EOF. This was simply done by least-square fitting a sine wave to each local EOF
of interest. The
instantaneous frequency can also be obtained from a complex wavelet
transform \cite{Farge}, by using information on the phase of the transform.

Our analysis does not reveal any evidence of self-similarity or fractality, as
obtained for the
previous synthetic data sets. Instead, we find a
preferred scale of variability between 3 and 5 years (not shown), which
corresponds to ENSO's low-frequency mode \cite{Rasmusson,jiangetal-1995}?
The first two local EOFs are consistently paired and in phase quadrature, which shows
that the nonlinear oscillation associated with this mode is robust and persists
throughout the sixty odd years being examined.

The computation of the instantaneous frequency allows us to detect an
abrupt frequency shift of the ENSO mode near 1960
(Fig.~\ref{fig-msssa-soi}). The characteristic periodicity goes
from 57 months (between 1943 and 1961) to 39 months (between
1963 and 1980). A comparable decrease in period was observed in the early 
1960s by Moron et al. \cite{moronetal-1998} in Tropical Pacific sea-surface
temperatures, by using multi-channel global SSA, and by Wang and
Wang \cite{wangwang-1996} in a sea-level pressure record at Darwin, using a wavelet
transform. 

\begin{figure}[h]
\caption{Instantaneous SOI frequencies, based on EOF-1 of MS-SSA at scale $W=128$.}
\label{fig-msssa-soi}
\end{figure}

Our MS-SSA approach improves on these previous results in two respects.
First, compared to both the global SSA analysis \cite{moronetal-1998}
and the wavelet and waveform (matching pursuit) approach \cite{wangwang-1996},
the combination of our local
SSA in a sliding window with the local sine-wave fit gives a
much sharper signature of the increase in instantaneous frequency. 
Second, we can check this regime change at 
different scales by performing the same analysis separately at each scale. 
In contrast, the wavelet
and waveform approaches need all the scales together in order to be able
to detect the frequency change.
We thus have at our disposal an additional degree of freedom to test for the 
reliability and robustness of the signal.

We find that, for $W= 64$ and $32$, the transition is not as sharp as with
$W=128$ (the value used in Fig.~19), but is still identifiable. 
This degradation is expected since the maximum period $M=W/3$ that can in principle be 
retrieved at these values of $W$
is less that the ENSO period of interest. This means in practice that only scales of 
the order of $W=128$ or larger will capture the ENSO period in question and the change
in it. At scales that are even larger, however, the analysis becomes limited in its
reliability by the finite length of the time series. Hence the value $W=128$ 
provides a good compromise between
the two conflicting constraints of resolution and statistical significance
\cite{vg-1989,vyg-1992,ghilyiou-1996,ghiltaricco-1997}.

\section{Concluding remarks}
\label{sec-conclusion}

We have presented a multi-scale extension of the orthogonal component
decomposition of a time series known as singular spectrum
analysis (SSA). This multi-scale extension, which we called MS-SSA,
has been performed by varying the width $W$ of the analyzing window, by analogy with
wavelet analysis. The successive analyses used a fixed
ratio -- equal to $2$ or $3$ in the numerical examples -- between the width $W$
and the order $M$ of the local covariance matrix. 

Our multi-scale SSA (MS-SSA), or data-adaptive wavelet analysis, extends
various other approaches in the same spirit. Mallat et
al. \cite{mallatetal-1997} describe a hierarchical binary tree of
basis functions for locally stationary time series. Lilly and Park
\cite{lillypark-1995} select their Slepian wavelets \cite{Slepian}
at a given scale by optimizing the
spectral energy in a given frequency band \cite{thomson-1982} and vary the scale.
Mann and Park \cite{parkmann-1997} have applied this
approach to a multivariate analysis of interannual temperature
variations. Wavelet packets
\cite{wickerhauser-1991,learnerwillsky-1995} can be defined as smooth
versions of the Walsh system used in binary signal processing. While they
extend wavelet analysis and take a step in the direction of satisfying
our requirements (A) and (B) above, the wavelet-packet approach does
not allow the same degree of flexibility and data-adaptivity as provided
by our method.  

Coifman and Saito \cite{coifmansaito-1997}
use the Karhunen-Lo\`eve expansion that underlies SSA to construct a
dictionary of orthonormal bases that has a tree structure.  Their
starting point is thus very similar to ours; Coifman and Saito, 
however, use additional criteria to choose
the orthonormal bases from the overdetermined dictionary in order to
perform a multi-scale analysis. Our Karhunen-Lo\`eve decomposition is
performed in moving windows of size $W$ that maintain a fixed
ratio to the order $M$ of the decomposition. As a consequence, our
orthogonal basis is unique, minimal and offers a very intuitive
interpretation of the decomposition. 

The empirical orthogonal functions (EOFs) play the same role in MS-SSA as the ``mother
wavelets.''  In contrast to standard wavelet transforms, our basis functions 
are data-adaptive, {\it i.e.} they are determined from a
diagonalization of the covariance matrix $C_M$ of the data within the observing window.
The data-adaptive character of the MS-SSA basis allows us to test
further for self-similar properties of a time series by comparing the
EOFs at different scales. 

In MS-SSA's orthogonal decomposition, the sum of a finite set of 
coefficients weighted by the corresponding ``wavelets,'' {\it i.e.}, the reconstructed
components (RCs) \emph{at the same scale}, reconstructs exactly the initial
signal. This is in contrast to the standard wavelet decomposition in
which one needs to sum over all scales or sum over an infinite set of
wavelets (for instance corresponding to the successive derivatives of
a Gaussian) at a given scale. We also propose an approximate 
``resolution of identity'' inversion formula to reconstruct the signal from
a nonorthogonal sum over scales of RCs that correspond
to a fixed EOF order. This formula
is the counterpart of the usual wavelet inversion formula and arises from the 
theory of ``frames'' \cite{daub-1992}.

We tested MS-SSA on a few irregular time series for which
conventional single-scale spectral analysis techniques would fail to reveal
their properties: Cantor sets and
intermittently amplifying multiplicative noise.  In many respects,
wavelets and MS-SSA behave in similar ways\,: MS-SSA decomposition onto the
first and second EOFs is very similar to a wavelet analysis that uses the
Gaussian mother wavelet and its
first derivative. But in the case of the approximate Cantor set $\tilde P_1$, the MS-SSA
decomposition gives a clearer insight on the way the set was
generated. In addition, the shape of the EOFs also allowed us to check
precisely for the self-similarity of that set.

The latter property is probably the most interesting one to investigate
in the future, as the change in shape of the EOFs as a function of scales
provide a new method to test for discrete scale invariance.
Indeed, the detection of the preferred scaling
ratio in a discretely scale-invariant system may be obtained by
finding the ratio of the multi-scale analysis that minimizes
the difference between the data-adaptive EOF ``wavelets'' obtained at
different scales. This test appears to be more robust than
previously used methods \cite{sornette-1998} and 
has been demonstrated successfully on a time
series that presents the scaling structure of the triadic Cantor set.

Finally, we applied the MS-SSA methodology to a real climatic time series,
given by the monthly Southern Oscillation index (SOI) for 1933-1996. The
SOI is a broadly used indicator for the El Ni\~no/Southern Oscillation (ENSO).
The sharp transition in ENSO's low-frequency mode from a period of roughly $5$ 
years to roughly $3$ years in the early 1960s refines earlier results obtained
using global SSA \cite{moronetal-1998} and standard wavelet \cite{wangwang-1996} 
analysis. It supports the Devil's staircase theory of ENSO \cite{jinetal-1994}
in the following sense\,: a sudden shift in frequency can be reconciled most
easily with the idea that an interdecadal change in mean thermocline depth in the
Tropical Pacific \cite{GhilJiang98,Knutson} causes the coupled ocean-atmosphere system
there to ``jump'' from one broad step of the staircase to another. We hope
that similar physical inferences will be facilitated in other areas of the
geosciences to which MS-SSA might be applied.

\section{Acknowledgments}
It is a pleasure to thank A. Fournier, M. Mann, J.-F. Muzy and N. Saito for stimulating
conversations and references.  This work was supported by NSF grant ATM95-23787 (MG and
PY), a NATO travel grant (PY) and research funds of UCLA's Institute of Geophysics
and Planetary Physics (IGPP) during PY's leave from LSCE--CEA, and NSF
grants EAR96-15357 and EAR97-06488 (DS).  This is contribution No. XXXX
of LSCE and contribution No. YYYY of IGPP at UCLA.

\newpage

\section{Appendix: Construction of the synthetic time series}

\subsection{Approximate triadic Cantor set}

The time series $\tilde P_1$  was obtained by 
using an Iterated Function System (IFS) \cite{barnsley-1993} to approximate
a simple Cantor set. An IFS
is a contracting stochastic map constructed to converge towards a
self-similar attractor of zero measure. The example we study consists
in a time series whose characteristics mimic the triadic
Cantor set. To generate this time series, we use the following
one-dimensional IFS:
\begin{equation}
x_{n+1} = \left\{ \begin{array}{ll}
{1 \over 3} x_n   & \mbox{with probability }\frac{1}{2},\\
{1 \over 3} x_n + {2 \over 3} & \mbox{with probability $\frac{1}{2}$}.
		  \end{array}\right.
\label{eq-ifs}
\end{equation}

Starting from an arbitrary point $x_0\in [0,1]$, it is straightforward
to verify that this stochastic IFS map converges to a triadic Cantor
set on $[0,1]$, characterized by the fractal dimension of $\log 2/
\log 3$ \cite{barnsley-1993}. We iterated Eq.~(\ref{eq-ifs}) 10000 times
and discarded the first 1000 values of the set so generated in order
to obtain a set $\{x_i\}$ that has converged close to the Cantor set
target. 

We sorted these values, $x_i\in [0,1]$, into increasing order,
$0 \leq x_0 \leq ... \leq x_i \leq x_{i+1} \leq ... \leq x_N \leq 1$.
A time series $\tilde P_1$ with, say $N=1000$ equidistant points, is
generated by interpolating the set $\{x_i\}$ every $\Delta t=1/N$ and
scaling it to $N$:
\begin{equation}
\tilde P_{1j}=\left\{
\begin{array}{ll}
1 & \mbox{if } \exists x_i \in [j/N,(j+1)/N],\\
0 & \mbox{otherwise},
\end{array}
\right.
\end{equation}
with $0 \leq j \leq N$.  As a result of the way this time series is
generated with the stochastic IFS, it contains residual noise with
respect to a perfect Cantor set both in time and scale, because of the
interpolation process (Fig.~\ref{fig-P1}). This means that the
smallest scales are not always of unit length, due to incomplete
interval division.

\subsection{The exact triadic Cantor set}

The second time series $P_1$ was obtained recursively by iterating
the binary map:
\begin{eqnarray}
1 & \mapsto & 101, \label{rule1}\\
0 & \mapsto & 000~, \label{rule2}
\end{eqnarray}
on the initial list $s_0=(1)$. At the $L$-th iteration, the list $s_L$
contains $3^L$ bits which have an exact Cantor set structure. We
generated the time series $P_1$ with $L=6$ iterations, so that
$P_{1j}=s_L(j)$, $1\leq j\leq 3^L$. A ``noisy'' Cantor time series
could be obtained by assigning a probability of occurence between rule
(\ref{rule1}) and $1\mapsto 111$. This procedure emulates the
incomplete division scheme towards smaller scales that we observe in
$\tilde P_1$.

A large
variety of Cantor sets can be obtained by modifying the rules
(\ref{rule1}, \ref{rule2}) in a straightforward manner.  For instance,
we generated a multi-rule Cantor set by iterating rules (\ref{rule1},
\ref{rule2}) three times, then used the rules
\begin{eqnarray}
1 & \mapsto & 110,\label{rule3}\\
0 & \mapsto & 000.\label{rule4}\nonumber
\end{eqnarray}
three more times, in order to obtain a set ${\hat P}_1$ of 729 ($=3^6$) points.

The fractal dimension of a set generated after iterating
infinitely many arbitrary permutations of rules (\ref{rule3}) and
(\ref{rule1}) is still $\log 2/ \log 3$, but the geometry of the
resulting set obviously depends on the sequence in which the rules
were applied. This is a clear instance where the fractal dimension is
not enough to characterize the geometric structure 
\cite{Mandelbrot-1982,Blumenfeld,Solis}.
Other rules and combinations of rules can be
used to obtain other fractal dimensions and, for a given dimension,
distinct geometric structures of the resulting set. 

\subsection{Log-periodic time series}

The functional form of the log-periodic process $P_2$ is simply:
\begin{equation}
P_2(t)=A+B (t-t_c)^\alpha \, \left\{1+C \cos\left[2 \pi \log(t-t_c)/
\log(\lambda)\right]\right\}.
\label{eq-logper}
\end{equation}
Here, we take $A=2$, $B=-1$, $C=0.2$, $\alpha=1/2$ and $\lambda=2$
(see Fig.~\ref{fig-lopger}). The critical time is $t_c=100$; we
regularly sample 1000 time steps between 1 and $t_c$ with a time
interval $\Delta t=0.1$, and define the $P_2$ time series by
$P_{2i}=P_2(i\Delta t)$, $1\leq i\leq 1000$. Note that in our case
$\alpha=1/2$ and thus $\lim_{t\to t_c} P_2(t)=A$; hence $P_2$ is indeed
continuous on the $[0,t_c]$ interval, although it is
not differentiable at $t=t_c$.

\subsection{The statistically self-similar process $P_3$}

The process $P_3$ is defined by
\begin{equation}
P_{3,i+1} = a_i P_{3i} + b_i,
\label{mapmap}
\end{equation}
where $a_i$ is a uniformly distributed random variable that can take
values larger than 1 --- here, $a_i \in [0.48,1.48]$ --- and $b_i$ is
uniformly distributed over $[0,1]$. In order to ensure that $P_3$ does
not grow to infinity but is ``globally'' stationary, we enforce the
technical condition that the average growth rate $\langle \ln a_i
\rangle$ be negative.  For the parameters taken above, it follows that
$\langle \ln a_i \rangle = -0.06747$. 

The apparently innocuous
self-affine map (\ref{mapmap}) leads to time series with surprisingly
rich properties \cite{sornettecont-1997,sornetteknopoff-1998}.  The
probability density function of $P$ has an asymptotic tail for large $P$ 
of the form of a power law $p(P) \sim P^{-(1 + \mu)}$, where the exponent $\mu$ is the 
smallest real solution of $\langle [a_i]^{\mu} \rangle = 1$. Complex
solutions $\mu$ of this probabilistic equation
also exist that have equal or larger real parts; they correspond to higher-order
corrections to the leading power-law behavior of $p(P)$ with 
log-periodic modulations. In practice, however, these modulations are not always
observable due to their averaging by the large fluctuations
in the $a_i$ and $b_i$ values \cite{Jogi}. 

In the present numerical example, the smallest real exponent is
$\mu \approx 1.5$. Since $\mu > 1$, the
mean $\langle P_{3i}\rangle$ is well-defined and finite. However,
since $\mu < 2$, the variance $\langle \left[ P_{3i}-\langle
P_{3i}\rangle\right]^2\rangle$ is infinite. In practice, this means
that, for a given realization and a fixed time length $T$, the
estimation of the variance by integration over $[t_0,t_0+T]$ is very
unstable as a function of the position $t_0$ of the time window and its
length $T$.  Furthermore, the variance diverges approximately as $T^{2-\mu}$
with the length $T$ of the time series.

\pagebreak


\newpage
\section{Figure captions}
\begin{description}
\item[Figure \ref{fig-P1}]: Iterated funtion system (IFS) simulation of a 
classical triadic Cantor set. (a) 1000-point interpolation 
to generate the $\tilde P_1$ time series; (b) blow-up
of $\tilde P_1$ between 1 and 333 in order to visualize the
(incomplete) triadic structure of the time series.

\item[Figure \ref{fig-lopger}]: Log-periodic time series $P_2$, with
critical point $t_c=100$, regularly sampled with 1000 points; see text and Eq.~(\ref{eq-logper})
for parameter values.

\item[Figure \ref{fig-p3}]: Multiplicative noise
process $P_3$: (a) individual realization over 512 equidistant points;
(b) histogram of the variations
of $P_3$, with 100 bins --- the ordinate represents the range of values (same
as in panel a) and the abscissa is a frequency count.

\item[Figure \ref{fig-soi}]: Variations of the Southern Oscillation
Index (SOI) between January 1933 and December 1996 (from the Climate Research Unit
Data at the University of East Anglia, U.K.). Time on the
abscissa in years and SOI on the ordinate centered to have mean zero and
normalized by its standard deviation.

\item[Figure \ref{fig-mcssa-p1}]: ``Global'' Monte-Carlo SSA of $\tilde
P_1$, with window width ({\it i.e.} number of lags)
 $M=40$. The vertical bars are 95\% confidence
intervals for 100 red-noise realizations with the same variance and
exponential-decorrelation time as $\tilde P_1(t)$. The diamonds
represent the singular values of the signal; those that stand out
above the error bars are statistically significant. The frequencies
associated with the latter are obtained by least-square fitting the
corresponding EOF to a sine function.

\item[Figure \ref{fig-rcssa-p1}]: Reconstructed components (RCs) of
$\tilde P_1$ with global SSA. (a--c) Reconstructions using groups of
eigenelements, as indicated in the legend of each panel; (d) the sum of
RCs 1--6. The SSA window is $M=40$, as in Fig.~\ref{fig-mcssa-p1}.

\item[Figure \ref{fig-wavt-P1}]: Wavelet transforms of $\tilde P_1$,
with scale increments of 3. The left panels use the Gaussian analyzing
function $\psi(x)$ (shown as ``Psi'' in the legend) and the right ones
use its derivative $\psi'(x)$ (shown as ``dPsi''); the scales
$a=3,9,27$ and 81 (for $n=1$--4, see text) are also indicated from top
to bottom on the left.

\item[Figure \ref{fig-msssa1-p1}]: Multi-scale SSA (MS-SSA) of
$\tilde P_1$, with scale increments of 3. The left
panels show the reconstruction of ${\tilde P}_1$ using the local
RCs associated with  EOF-1, the right panels that using the RCs
obtained by projection onto the local EOF-2; the window widths $W = 9, 27, 81$ and
$243$ are shown from top to bottom on the left (like the scale $a$ in Fig.~7).

\item[Figure \ref{fig-eof-p1}]: Normalized EOFs 1--4 (panels a--d) of
$\tilde P_1$ at $b=333$. The EOFs at scale $3^n$
were multiplied by a factor of ${3}^{(n-1)/2}$; the abscissa in each
panel is linearly normalized to $[0,1]$. The scales are $W=9$, 27, 81 and 243,
respectively in thick continuous, thin continuous, dotted and dashed
lines (see also legend in panel a).

\item[Figure \ref{fig-eof-p1-2k}]: Normalized EOFs 1--4 of
$\tilde P_1$ at $b=333$. The EOFs at scale $2^n$
are multiplied by a factor of ${2}^{(n-1)/2}$; abscissa normalized as
in Fig.~\ref{fig-eof-p1}. The scales are $W=16$, 32, 64, 128 and
256, respectively in thick continuous, thin continuous, dotted, dashed
and dot-dashed lines (see also legend in panel b).

\item[Figure \ref{fig-msssa-p1mlt}]: MS-SSA of
$\hat P_1$, with scale increments of
3. The left panels contain the RCs associated with EOF-1, the right
panels those associated with EOF-2; the same window widths $W$ 
as in Fig. \ref{fig-msssa1-p1} (shown on the left).


\item[Figure \ref{fig-inversion-p1}]: Reconstruction of $P_1$ using 
formula (\ref{dsfgsg}) for scales that are powers of 3.
The top panel shows the time series $P_1$ derived from the exact triadic Cantor set.
The remaining panels show the reconstructions of $P_1$ using the local RCs based
on EOF-2. The second panel uses RC-2 for the single scale $N=9$; this
corresponds to Eq. (\ref{dsfgsg}) with a single value of $m=2$, $a=3$ and thus
$W_2 = 9$. The third panel shows the sum (\ref{dsfgsg}) with two terms, corresponding to 
$W_2 = 9$ and $W_3 = 27$;
the fourth panel shows the sum (\ref{dsfgsg}) with three terms, 
corresponding to  $W_2 = 9$, $W_3 = 27$ and $W_4 = 81$;
the bottom panel shows the sum (\ref{dsfgsg}) with four terms, corresponding to 
$W_2 = 9$, $W_3 = 27$, $W_4 = 81$ and $W_5 = 243$ and still the same order $k=2$ of the EOF.

\item[Figure \ref{fig-eof-p2}]: Normalized EOFs 1--4 (panels a-d) of $P_2$ at
$t=t_c-W/2$, so that each window ends at $t_c$. The EOFs at scale
$2^n$ are multiplied by a factor of ${2}^{(n-1)/2}$, and the abscissa
is normalized to $[0,1]$. The scales are $W=16$, 32, 64, 128 and 256,
respectively in thick continuous, thin continuous, dotted, dashed and
dash-dotted lines (see panel a).

\item[Figure \ref{fig-msssa-p2}]: MS-SSA of $P_2$ with RC-1 (left column)
and RC-2 (right column). The scales $W$ vary from 16 to 256, going from the 
top to the bottom (indicated along the left edge of the figure).



\item[Figure \ref{fig-rcssa-p3}]: RCs of $P_3$ using
global SSA. Panels (a--d) represent reconstructions that use groups of RCs as
indicated to the left of each panel; panel e contains the sum of RCs 1-8.

\item[Figure~\ref{fig-wavlt-p3}]: Wavelet analysis of $P_3$ with a
Gaussian wavelet and its derivative. The scales vary from $a=2$ to
$32$ in powers of $2$; same layout as for Fig. 7.

\item[Figure~\ref{fig-msssa-p3}]: MS-SSA of $P_3$ with RC-1
and RC-2. The scales $W$ vary from 4 to 64, in powers of $2$;
same layout as in Fig. 8.


\item[Figure~\ref{fig-mesheof-p3}]: Evolution in time of the local EOF-1's for
$P_3$, at $W=32$ ($M=W/3 = 10$). The axes are the time and the lag.

\item[Figure~\ref{fig-msssa-soi}]: Instantaneous frequencies of the
local EOFs 1 (solid) and 2 (dotted) for the SOI time series in Fig. 4; 
the MS-SSA scale is $W=128$ and the number of lags is $M=W/3=42$.
\end{description}


\begin{thebibliography}{10}

\bibitem{ghil-childress} M.~Ghil and S.~Childress.  \newblock {\em
Topics in Geophysical Fluid Dynamics: Atmospheric Dynamics, Dynamo
Theory and Climate Dynamics}.  \newblock Springer-Verlag, New York,
1987.

\bibitem{Mandelbrot-1982} B.~B.  Mandelbrot.  \newblock {\em The
Fractal Geometry of Nature}.  \newblock Freeman, San Francisco, 2nd
edition, 1982.

\bibitem{kagan-1994} Y.~Y.  Kagan.  \newblock Observational evidence
for earthquakes as a nonlinear dynamic process.  \newblock {\em
Physica D}, 77:160--192, 1994.

\bibitem{peltier-1996b} W.~R. Peltier.  \newblock Phase transition
modulated mixing in the mantle of the earth.  \newblock {\em
Phil. Trans.  Roy. Soc. London Series A}, 354:1425--1443, 1996.

\bibitem{rhodesanderson-1996} C.~J. Rhodes and R.~M.  Anderson.
\newblock Power laws governing epidemics in isolated populations.
\newblock {\em Nature}, 381:600--602, 1996.

\bibitem{gould-1996} S.~J. Gould.  \newblock {\em Full House}.
\newblock Harmony Books, New York, 1996.

\bibitem{hannan-1960} E. J. Hannan. \newblock {\em Time Series
Analysis}. \newblock Methuen, New York, 1960.

\bibitem{percivalwalden-1993} D.~B. Percival and A.~T.  Walden.
\newblock {\em Spectral Analysis for Physical Applications}.
\newblock Cambridge University Press, Cambridge, UK, 1993.

\bibitem{BK-1986} D.~S. Broomhead and G.~P. King.  \newblock
Extracting qualitative dynamics from experimental data.  \newblock
{\em Physica D}, 20:217--236, 1986.

\bibitem{vg-1989} R.~Vautard and M.~Ghil.  \newblock Singular spectrum
analysis in nonlinear dynamics, with applications to paleoclimatic
time series.  \newblock {\em Physica D}, 35:395--424, 1989.

\bibitem{vyg-1992} R.~Vautard, P.~Yiou, and M.~Ghil.  \newblock
Singular spectrum analysis: A toolkit for short noisy chaotic signals.
\newblock {\em Physica D}, 58:95--126, 1992.

\bibitem{ghilvaut-1991} M. Ghil and R. Vautard. \newblock Interdecadal
oscillations and the warming trend in global temperature time
series. \newblock {\em Nature} 350:324--327, 1991.

\bibitem{yiouetal-1995b} P.~Yiou, M.~F. Loutre, and E.~Baert.
\newblock Spectral analysis of climate data.  \newblock {\em Surveys
Geophys.}, 17:619--663, 1996.

\bibitem{moronetal-1998} V. Moron, R. Vautard, and M. Ghil. \newblock
Trends, interdecadal and interannual oscillations in global
sea-surface temperatures. \newblock {\em Clim. Dyn.}, 14:545--569, 1998.

\bibitem{meyer-1989} Y.~Meyer.  \newblock {\em Ondelettes et
Op\'erateurs {I}: Ondelettes}.  \newblock Hermann, Paris, 1989.

\bibitem{daub-1992} I.~Daubechies.  \newblock {\em Ten Lectures on
Wavelets}.  \newblock SIAM, Philadelphia, PA, 1992.

\bibitem{meyer-1993} Y.~Meyer.  \newblock {\em Wavelets: Algorithms
and Applications}.  \newblock SIAM, Philadelphia, PA, 1993.

\bibitem{arneodoetal-1993} A.~Arneodo, F.~Argoul, J.~F. Muzy, and
M.~Tabard.  \newblock Beyond classical multifractal analysis using
wavelets: {U}ncovering a multiplicative process hidden in the
geometrical complexity of diffusion limited aggregates.  \newblock
{\em Fractals}, 1:629--649, 1993.

\bibitem{holschneider-1995} M.~Holschneider.  \newblock {\em Wavelets:
An Analysis Tool}.  \newblock Clarendon Press, New York, 1995.

\bibitem{Courant} R.~Courant and D.~Hilbert. 
\newblock {\em Methods of Mathematical Physics.}
\newblock New York, Interscience Publishers, Vol. I,  1953.

\bibitem{dettingeretal-1994} M. D. Dettinger, M. Ghil, C. M. Strong,
W. Weibel, and P. Yiou. \newblock Software expedites singular-spectrum
analysis of noisy time series. \newblock {\em Eos Trans. AGU}, 76:12,
20, 21, 1995.

\bibitem{ghilyiou-1996} M. Ghil and P. Yiou. \newblock Spectral
methods: {W}hat they can and cannot do for climatic time
series. \newblock In {\em Decadal Climate Variability: Dynamics and
Predictability}, D. Anderson and J. Willebrand, editors, Elsevier,
Amsterdam, pp. 446-482, 1996.

\bibitem{ghiltaricco-1997} M. Ghil and C. Taricco. \newblock Advanced
spectral analysis methods. \newblock In {\em Past and Present
Variability of the Solar-Terrestrial System: Measurement, Data
Analysis and Theoretical Models}, G. Cini Castagnoli and A. Provenzale
(eds.), Societa Italiana di Fisica, Bologna, \& IOS Press, Amsterdam,
pp. 137--159, 1997.

\bibitem{yiou-1994} P. Yiou. \newblock Dynamique du Pal\'eoclimat: Des
Donn\'ees et des Mod\`eles. \newblock Ph.D. thesis, Universit\'e
Pierre et Marie Curie, Paris 6, 1994.

\bibitem{Gibsonetal}  J.F.~Gibson, J.D. Farmer, M. Casdagli and S. Eubank, An 
analytic approach to practical space reconstruction. {\em Physica D}, 57:1--30, 1992.

\bibitem{KeppenneGhil92} C.L.~Keppenne and M.~Ghil.
\newblock Adaptive filtering and prediction of the Southern Oscillation index.
\newblock {\em J. Geophys. Res.}, 97:20449--20454, 1992.

\bibitem{GhilJiang98} M.~Ghil and N.~Jiang.
\newblock Recent forecast skill for the El Niño/Southern Oscillation. 
\newblock {\em Geophys. Res. Lett.}, 25:171--174, 1998.

\bibitem{dubrulleetal-1997} B.~Dubrulle, F.~Graner, and D.~Sornette,
editors.  \newblock {\em Scale Invariance and Beyond}. EDP Sciences
and Springer, Berlin, 1997. 

\bibitem{barnsley-1993} M.~F. Barnsley.  \newblock {\em Fractals
Everywhere}.  \newblock Academic Press, Boston, 2nd edition, 1993.

\bibitem{sornette-1998} D.~Sornette.  \newblock Discrete scale
invariance and complex dimensions.  \newblock {\em Physics Reports},
297:239--270, 1998.

\bibitem{Jona} G. Jona-Lasinio. The renormalization group: a probabilistic view.
{\em Nuovo Cimento}, 26B:99, 1975.

\bibitem{Nauen} M. Nauenberg. Scaling representations for critical
phenomena. {\em J.Phys. A}, 8:925, 1975.

\bibitem{Nieme} Th. Niemeijer and J.M.J. van Leeuwen. {\em In Phase Transitions and
Critical Phenomena}, Vol.6, C. Domb and M.S. Green, eds. (Academic Press, London,
1976), p.425. 

\bibitem{DIL} B. Derrida, L. De Seze and C. Itzykson. Fractal structure of zeros
in hierarchical models. {\em J. Stat. Phys.}, 33:559-569, 1983;
B. Derrida, C. Itzykson and J.M. Luck. 
Oscillatory critical amplitudes in hierarchical models.
{\em Commun.Math.Phys.}, 94:115-132, 1984.

\bibitem{Fournier} D. Bessis, J.-D. Fournier, G. Servizi, G. Tourchetti and S.
Vaienti. Mellin transforms of correlation integrals and generalized dimension of
strange sets. {\em Phys. Rev. A}, 36:920--928, 1987.

\bibitem{crash} D.~Sornette and A.~Johansen.
Large financial crashes. {\em Physica A}, 245:411--422, 1997.

\bibitem{ghilmulh-1985} D. Dee, and M. Ghil.
\newblock  Boolean difference equations, I:
Formulation and dynamic behavior.
\newblock {\em SIAM J. Appl. Math.}, 44:111--126, l984;
M. Ghil and A. Mullhaupt. \newblock Boolean
delay equations {II}: Periodic and aperiodic solutions. \newblock {\em
J. Stat. Phys.}, 41:125--173, 1985.

\bibitem{Ghiletal87paleo} M.~Ghil,  A. Mullhaupt and P. Pestiaux,  Deep water formation
and Quaternary glaciations, {\em Clim. Dyn.}, 2:1--10, 1987.

\bibitem{Wrightetal90paleo} D.G.~Wright, T. F. Stocker, and L. A. Mysak.
\newblock A note on Quaternary climate modelling using Boolean delay equations.
\newblock {\em Clim. Dyn.}, 4:263--267, 1990.

\bibitem{DarbyMysak93} M.S.~Darby and L.A.~Mysak, A Boolean delay equation model
of an interdecadal arctic climate cycle, {\em Clim. Dyn.}, 8:241--246, 1993.

\bibitem{Saunders} A.~Saunders, A., A Boolean Delay Equation Model of ENSO Variability,
M. S. Thesis, UCLA (1998).

\bibitem{sornettecont-1997} D.~Sornette and R.~Cont.  \newblock
Convergent multiplicative processes repelled from zero: power laws and
truncated power laws.  \newblock {\em J. Phys. I France}, 7: 431--444,
1997.

\bibitem{sornetteknopoff-1998} D.~Sornette. \newblock Linear
stochastic dynamics with nonlinear fractal properties.  \newblock {\em
Physica A}, 250:295--314, 1998.  

\bibitem{ghiletal-1991} M.~Ghil, M.~Kimoto, and J.~D.  Neelin.
\newblock Nonlinear dynamics and predictability in the atmospheric
sciences.  \newblock {\em Rev. Geophys.}, 36 (Supplement 
U. S. National Report IUGG, 1987--91):46--55, 1991.  

\bibitem{Rasmusson} E.M.~Rasmusson, X.~Wang, and C.F.~Ropelewski.
\newblock The biennial component of ENSO variability.
\newblock {\em J. Mar. Syst.}, 1: 71--96, 1990.

\bibitem{jiangetal-1995} N.~ Jiang, D.~Neelin, and M.~Ghil. \newblock
Quasi-quadrennial and quasi-biennial variability in the equatorial
{P}acific. \newblock {\em Clim. Dyn.}, 12:101--112, 1995.

\bibitem{jinetal-1994} P. Chang, B. Wang, T. Li, and L. Ji.
\newblock Interactions between the
seasonal cycle and the Southern Oscillation: Frequency entrainment
and chaos in an intermediate coupled ocean-atmosphere model.
{\em Geophys. Res. Lett.}, 21:2817--2820, 1994; 
F.-F. Jin, J.~D.  Neelin, and M.~Ghil.
\newblock El {N}i\~no on the {D}evil's staircase: Annual subharmonic
steps to chaos.  \newblock {\em Science}, 264:70--72, 1994;
E. Tziperman, L. Stone, M. Cane and H. Jarosh. \newblock  El Ni\~no
chaos:  Overlapping of resonances between the seasonal cycle
and the Pacific ocean-atmosphere oscillator. \newblock {\em Science}, 264:72--74, 1994.

\bibitem{wangwang-1996} B. Wang and Y. Wang. \newblock Temporal
structure of the Southern Oscillation as revealed by waveform and
wavelet analysis. \newblock {\em J. Clim.}, 9:1586--1598, 1996.

\bibitem{allensmith-1995} M.~R. Allen and L.~A. Smith.  \newblock
Monte {C}arlo {SSA}: Detecting irregular oscillations in the presence
of coloured noise.  \newblock {\em J. Clim.}, 9:3373--3404, 1996.

\bibitem{childers-1978} D. G. Childers, editor. {\em Modern Spectrum
Analysis}. \newblock IEEE Press, New York, 1978.

\bibitem{thomson-1982} D. J. Thomson, Spectrum estimation and harmonic
analysis. \newblock {\em IEEE Proc.}, 70:1055--1096, 1982.

\bibitem{Feller} W.~Feller.
\newblock {\em An Introduction to Probability Theory and its Applications}.
\newblock John Wiley and Sons, New York, 1971.

\bibitem{ghilrizzoli-1991} Ghil, M. and P. Malanotte-Rizzoli. \newblock
Data assimilation in meteorology and oceanography. \newblock {\em
Adv. Geophys.}, 33:141--266, 1991; Miller, R. N., M. Ghil and
F. Gauthiez. \newblock Advanced data assimilation in strongly
nonlinear dynamical systems. \newblock {\em J. Atmos. Sci.},
51:1037--1056, 1994.

\bibitem{SSearthquake} D.~Sornette and C.G.Sammis. Complex critical
exponents from renormalization group theory of earthquakes\,: Implications for earthquake
predictions. {\em J.Phys.I France}, 5:607--619, 1995.

\bibitem{Johansen} A.~Johansen, D.~Sornette, H.~Wakita, U.~Tsunogai, W.I.~Newman and H.~Saleur.
\newblock Discrete scaling in earthquake precursory phenomena\,: evidence in the Kobe earthquake,
Japan. \newblock {\em J.Phys.I France}, 6:1391--1402, 1996.

\bibitem{Farge} M.~Farge.
\newblock Wavelet transforms and their applications to  turbulence.
\newblock {\em Ann. Rev. Fluid Mech.}, 24:395--457, 1992.

\bibitem{mallatetal-1997} S.~Mallat, G.~Papanicolaou, and Z.~Zhang.
\newblock Adaptive covariance estimation of locally stationary
processes.  \newblock {\em Annal. of Statistics}, 26:1--47, 1998. 

\bibitem{lillypark-1995} J.~M. Lilly and J.~Park.  \newblock
Multiwavelet spectral and polarization analyses of seismic records.
\newblock {\em Geophys. J. Int.}, 122:1001--1021, 1995.

\bibitem{Slepian} S. Slepian. Prolate spheroidal wave functions, Fourier
analysis and uncertainty-V: The discrete case.  {\em Bell. Syst. Tech.
J.}, 1371--1430, 1978.

\bibitem{parkmann-1997} J.~Park and M.~E. Mann.  \newblock Interannual
temperature events and shifts in global temperature: {A}
``multiwavelet'' correlation approach.  \newblock Preprint, 1997.

\bibitem{wickerhauser-1991} M.~V. Wickerhauser.  \newblock Lectures on
Wavelet Packet Algorithms.  \newblock Technical Report, Dept.
Mathematics, Washington U., St. Louis, MO, 1991.

\bibitem{learnerwillsky-1995} R.~E. Learned and A.~S.  Willsky.
\newblock A wavelet packet approach to transient signal
classification.  \newblock {\em Appl. Comput. Harmonic Anal.},
2:265--278, 1995.

\bibitem{coifmansaito-1997} R.~R. Coifman and N.~Saito.  \newblock The
local {K}arhunen-{L}o\`eve bases.  
\newblock  {\em Proc. IEEE Intl. Symp. Time-Frequency and Time-Scale Analysis}, 
pp. 129--132, 1996.

\bibitem{Blumenfeld} R.~Blumenfeld and B.B. Mandelbrot. 
L\'evy dusts, Mittag-Leffler statistics, mass fractal lacunarity, and
perceived dimension. {\em Phys. Rev. E}, 56:112--118, 1997.

\bibitem{Solis} F.J.~Solis and L. Tao. Lacunarity of random fractals.
{\em Phys. Lett. A}, 228:351--356, 1997.

\bibitem{Jogi} P.~J\"ogi, D.~Sornette and M.~Blank.
Fine structure and complex exponents in power law distributions from random maps.
{\em Phys. Rev. E}, 57:120--134, 1998.

\bibitem{Knutson} T.R.~Knutson, S. Manabe, and D. Gu. Simulated
ENSO in a global coupled ocean-atmosphere model:
Multidecadal amplitude modulation and $CO_2$ sensitivity.
{\em J. Clim.}, 10:138-161, 1997; D. Gu, and S. G. H. Philander.
Secular changes of annual and interannual variability in the
tropics during the past century. {\em J. Clim.}, 8:64-876, 1995.






\end{thebibliography}
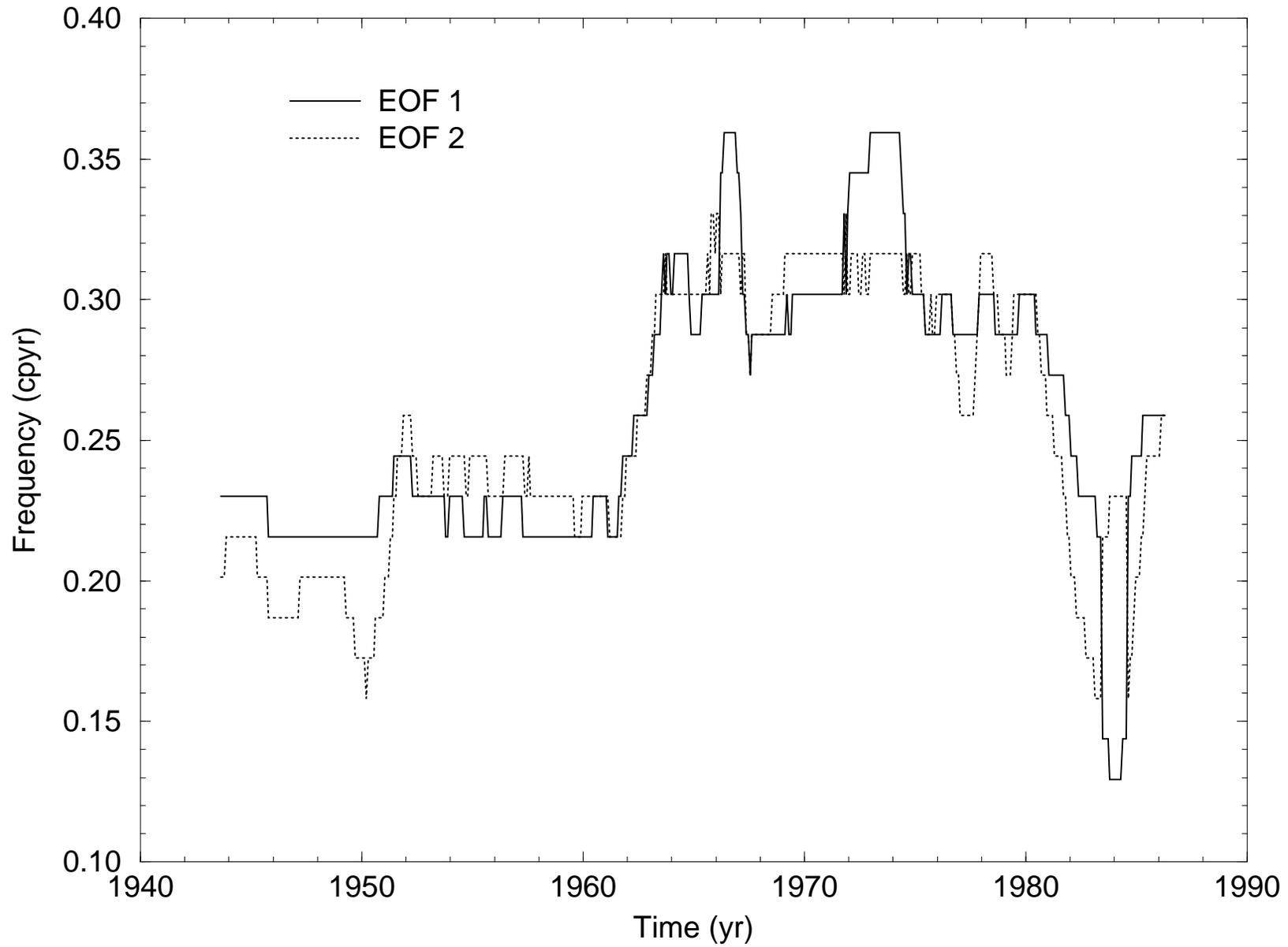
\end{document}